\documentclass[a4paper,11pt]{article}
\usepackage{jheppub} % for details on the use of the package, please see the JINST-author-manual
\usepackage{lineno}
\usepackage{xspace}
\usepackage{xcolor}
\usepackage{comment}
\usepackage{subfigure}
\usepackage{mathrsfs}
\newcommand{\jpsi} {\ensuremath{{\mathrm J}/\psi}\xspace}
\newcommand{\psip} {\ensuremath{\psi'}\xspace}
\newcommand{\rhozero} {\ensuremath{\rho^0}\xspace}
\newcommand{\pentaquark} {\ensuremath{P_C(4380)}\xspace}
\newcommand{\tetraquark} {\ensuremath{T_{cc\bar{c}\bar{c}}}\xspace}

\newcommand{\starlight}{STARlight\xspace}

\newcommand{\AxE}{\ensuremath{A\times{\epsilon}}}

\newcommand{\pp}           {pp\xspace}

\newcommand{\pA}           {\mbox{p--A}\xspace}
\newcommand{\AAcoll}           {\mbox{A--A}\xspace}

\newcommand{\pt}           {\ensuremath{p_{\rm T}}\xspace}

\newcommand{\nineH}        {$\sqrt{s}~=~0.9$~Te\kern-.1emV\xspace}
\newcommand{\seven}        {$\sqrt{s}~=~7$~Te\kern-.1emV\xspace}
\newcommand{\eight}        {$\sqrt{s}~=~8$~Te\kern-.1emV\xspace}
\newcommand{\twoH}         {$\sqrt{s}~=~0.2$~Te\kern-.1emV\xspace}
\newcommand{\twosevensix}  {$\sqrt{s}~=~2.76$~Te\kern-.1emV\xspace}
\newcommand{\five}         {$\sqrt{s}~=~5.02$~Te\kern-.1emV\xspace}
\newcommand{\fiveExactly}  {$\sqrt{s}~=~5$~Te\kern-.1emV\xspace}
\newcommand{\twosevensixnn}{$\sqrt{s_{\mathrm{NN}}}~=~2.76$~Te\kern-.1emV\xspace}
\newcommand{\fivenn}       {$\sqrt{s_{\mathrm{NN}}}~=~5.02$~Te\kern-.1emV\xspace}

\newcommand{\GeVc}         {Ge\kern-.1emV/$c$\xspace}
\newcommand{\MeVc}         {Me\kern-.1emV/$c$\xspace}
\newcommand{\TeV}          {Te\kern-.1emV\xspace}
\newcommand{\GeV}          {Ge\kern-.1emV\xspace}
\newcommand{\GeVtwo}       {Ge\kern-.1emV$^2$\xspace}
\newcommand{\MeV}          {Me\kern-.1emV\xspace}
\newcommand{\GeVmass}      {Ge\kern-.1emV/$c^2$\xspace}
\newcommand{\MeVmass}      {Me\kern-.1emV/$c^2$\xspace}

\title{\boldmath AI-powered full-data set search for new physics in ultraperipheral and diffractive events}

\author[1]{S. Ragoni\note{Corresponding author.}}
\author[]{, B. Kinkaid}
\author[]{, J. Seger}
\author[]{, C. Anson}
\author[]{, D. Tlusty}
\affiliation{Creighton University,\\2500 California Plz, Omaha, \\NE 68178, United States, USA}

\emailAdd{simone.ragoni@cern.ch}
\emailAdd{briannakinkaid@creighton.edu}
\emailAdd{jseger@creighton.edu}
\emailAdd{chrisanson5@gmail.com}
\emailAdd{tlusty@gmail.com}

\abstract{We present possible strategies for anomaly detection of rare particle decays and exotic hadrons, such as pentaquarks, in low-background environments such as those characteristic of diffractive events and ultraperipheral \pp, \pA, or \AAcoll collisions at the CERN Large Hadron Collider (LHC). Our models are trained with  toy samples representing the UPC processes measured until now by the ALICE Collaboration. When samples containing rare processes such as $\jpsi\rightarrow4\pi$ and pentaquark production, where the number of injected pentaquark events is estimated based on current experimentally available upper limits, and those for $\jpsi\rightarrow4\pi$ are estimated through the branching ratio of the decay channel, are analyzed, the rare processes are flagged as anomalous by the models. This approach demonstrates the applicability of such a technique for searches for new physics in the current and future data sets at collider experiments with high purity, while also allowing for the measurement of upper limits for the production of exotica. }

\begin{document}
\maketitle
\flushbottom

\section{Introduction}
Ultra-peripheral collisions (UPCs), and more generally, diffractive processes, have been studied extensively. 
%gained overwhelming popularity in recent years. 
Experiments such as ALICE \cite{alice-vector-meson}, ATLAS \cite{ATLAS:2022ryk}, CMS \cite{cms-vector-meson}, and LHCb \cite{lhcb-vector-meson} at the CERN Large Hadron Collider (LHC), and STAR \cite{star-vector-meson} at the Relativistic Heavy Ion Collider (RHIC), have published numerous results on several experimental signatures of exclusive photoproduction processes and exclusive lepton pair production. Results on exclusive \rhozero \cite{ALICE:2020ugp, STAR:2007elq}, \jpsi \cite{ALICE:2023jgu}, \psip \cite{Acharya:2748581, LHCb:2022ahs}, and so on, have been used to study both nuclear shadowing and gluon saturation.

More recently, ALICE and STAR have released measurements of excited \rhozero states \cite{ALICE:2024kjy, Gorbunov:2009ki}, opening the study of resonances whose properties are not well documented in the Particle Data Group (PDG) \cite{ParticleDataGroup:2024cfk}.
%which leads to the direction of poorly known or unknown resonances within the PDG. 
Similarly, the GlueX collaboration has published exclusion limits on the search for pentaquark states in photoproduction \cite{GlueX:2019mkq}. In addition, \cite{Goncalves:2021ytq} explores the possibility of measuring fully-charmed tetraquark production in UPCs, which may be produced in the photon-fusion channel. This could provide access to the wave function itself and answer questions about the tetraquark production mechanism.

These rare processes can involve decay channels that are not often explored. For example, the measurement of the pentaquark by LHCb in inclusive events \cite{LHCb:2019kea} is made in the \jpsi $+$ p channel, and tetraquark searches have occurred in the di-\jpsi channel \cite{LHCb:2014zwa}. 
%processes are very rare, or result in topologies that are not often explored in the respective fields. For example, the measurement of the pentaquark by LHCb in inclusive events \cite{LHCb:2019kea} in the \jpsi$+$p channel, and of resonances in the di-\jpsi channel. {\bf these first two sentences need editing} 
These searches require dedicated cut-based analyses, and may be limited by combinatorial backgrounds. 
%It is possible to leverage the strikingly low multiplicities commonly found in UPCs and diffractive events to implement more sophisticated searches for these resonances, than what is allowed in cut-based measurements which are limited by combinatorial backgrounds. 
While machine learning techniques have yet to be prominently used in the analysis of ultra-peripheral collisions, recent work has suggested that forms of machine learning could be an effective way to isolate UPC events without specific trigger selections \cite{Ragoni:2024jhg}. In addition, \cite{Ragoni:2024ovv} shows how anomaly detection could be a viable approach to implement automated searches for new processes and rare decays without relying on cut-based analyses, but does not yet fully explore the possibilities at colliders. 

Traditional searches at colliders usually rely on particle identification (PID) techniques combined with cut-based analyses, e.g. \cite{ALICE:2023jgu}, or supervised machine learning, e.g. \cite{ALICE:2022sco}. They all rely heavily on predefined decay topologies, or available Monte Carlo simulations, so that cuts can be optimized or models may be trained.

The approach presented here further expands on the proof-of-concept technique shown in \cite{Ragoni:2024ovv} to include the simulation of experimental PID capabilities. This allows us to fully explore an alternative to traditional data analysis techniques
%. This approach will be a viable technique to apply 
that may be applied to the data sets already available at STAR, the upcoming data sets from Run 3 and Run 4 at the ALICE experiment, and at the future Electron-Ion Collider (EIC). 

\section{Methodology}
The aim of this paper is to provide possible autoencoder \cite{Hinton:2006tev} architectures and input-vector strategies to enable current and future experiments, e.g. ALICE and ePIC at the EIC, to use anomaly detection to search for rare processes and new physics. We require samples for both training and validation that replicate data sets that have or will be collected by the experiments. We use the ALICE experiment as our primary test case, since it has measured the most resonances and processes among the currently operating experiments. We focus on processes above an invariant mass of 1.5~\GeVmass, since the mass region below 1.5~\GeVmass is dominated by coherent \rhozero production and non-resonant dipion production. In addition, since current experiments confirm that the production cross section for incoherent production is a few orders of magnitude lower than coherent production, we focus on coherent production. We therefore assume that $\pt\sim 0$~\GeVc for each resonance and process under consideration. We have modeled $\gamma\gamma\rightarrow\mu\mu$, which is the main contribution to the continuum. We have also modeled non-resonant dikaon production in two channels, $\gamma\gamma\rightarrow KK$ and $\gamma\mathcal{P} \rightarrow KK$\footnote{$\mathcal{P}$ demarks the pomeron.}, which are indistinguishable from each other in this context, as well as $\jpsi \rightarrow \mu^+\mu^-$, $\jpsi \rightarrow {\rm p}\bar{\rm p}$, $\psip \rightarrow \mu^+\mu^-$ and $\psip \rightarrow \mu^+\mu^-\pi^+\pi^-$. All processes have been generated to reproduce the kinematics found in  \starlight~\cite{Klein:2016yzr}, which is commonly used in photoproduction physics at high-energy colliders. All processes have been generated with isotropic distributions in the centre-of-mass frame of the original particle, or in the $\gamma\gamma$ frame, for convenience, since angular observables are quite susceptible to acceptance and detector efficiencies, and thus not suitable for large-scale automatic searches.

We have also generated processes that reproduce interesting new decay channels that could be measured with the statistics collected by experiments in future runs, and processes that have been extensively looked for in experiments, but whose observation is still outside of our grasp. We have focused on $\pentaquark \rightarrow \jpsi + \text{p}$, the pentaquark at a mass of 4.38~\GeVmass searched for by the GlueX collaboration \cite{GlueX:2019mkq}, and $\jpsi\rightarrow4\pi$, which is listed among the possible decay channels by the PDG \cite{ParticleDataGroup:2024cfk}, and can realistically be measured even with current data at e.g. ALICE.

All processes are generated with $\pt\sim 0$~\GeVc as stated above, and the proper kinematics is propagated to the decay particles. This is crucial in this effort, since the decay particles pass through a particle identification (PID) procedure. We have implemented the functional form of the ALICE TPC PID response, and assigned values for the energy loss $dE/dx$ signals to each of the decay daughters.
%, based on a $3\sigma$ band around the central value of the functional form. 
This is then used to measure the difference between the central line of the PID bands and the $dE/dx$ values obtained. 
%by the PID smearing. 
This is expressed in terms of $\sigma$, which corresponds to the standard deviation of the PID distribution. It is usually expressed in multiples of $\sigma$ as $N\sigma$ \cite{ALICE:2020jsh}.

\begin{figure}[b]
\includegraphics[width=1.\columnwidth]{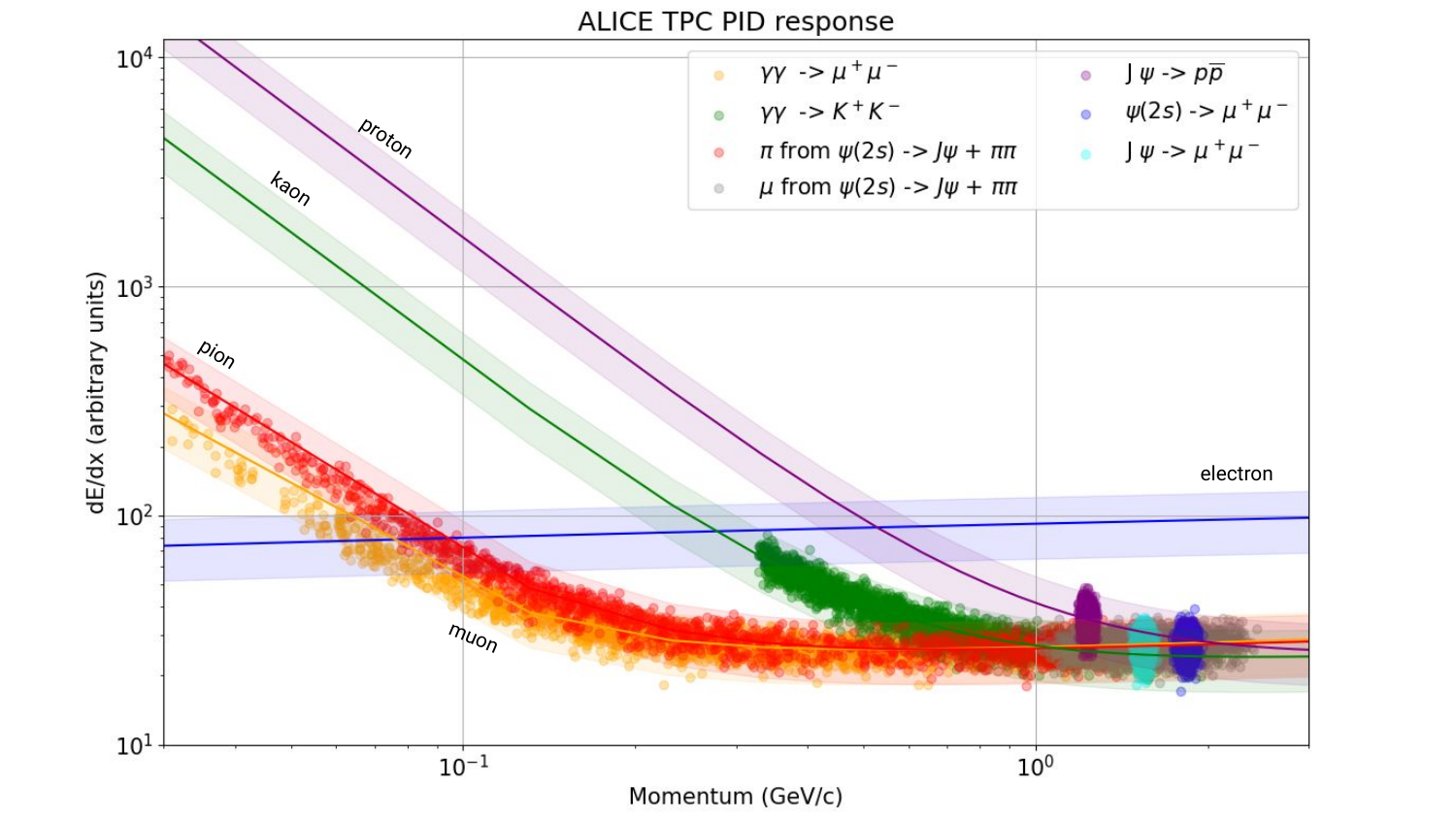}
\caption{\label{fig:tpc-response} The functional form of the ALICE TPC PID response is shown with the colored bands. Decay particles from the simulated processes, folded with a 3$\sigma$ PID response, are shown with dots. %to ensure the possibility to use the data for anomaly detection with identified particles.
}
\label{fig:pid}
\end{figure}

Hence, we are effectively able to model the experimental PID of the decay particles with a fast and computationally inexpensive generation, without relying on the simulation framework of the experiments. The functional forms of the TPC PID response \cite{ALICE:2014sbx} for a variety of particles in the $(dE/dx, p)$ plane, where $p$ indicates total momentum, are shown in Fig.~\ref{fig:tpc-response}. Colored bands mark the range of particles falling within $3\sigma$ of the expected $dE/dx$ for a given momentum. Decay particles from each of the included processes are shown as dots on the plot.  The overlap between the processes in the region of the resonances is of particular interest; this has driven the design choice for the implementation of the autoencoders, as shown below. 

%---

\subsection{Autoencoder architectures}
Two neural networks for autoencoders \cite{Hinton:2006tev} have been implemented using different strategies to be sensitive to rare processes \cite{Farina:2018fyg}, using the \textit{Keras} library \cite{chollet2015keras}. In this type of neural network, the model learns to represent the data using a lower-dimensional representation, also referred to as latent space, from which it can then reconstruct the original input. The autoencoder is made up of an encoder, that projects the input vector to the latent space, and a decoder, that reconstructs the representation of the vector from the latent space. The two autoencoders differ in the input vectors and in the number of dimensions for the latent space. In each case, the features associated with an event were standardized to ensure consistent input for the machine learning model. Both autoencoders were trained with a sample containing 30,000 events each from the various processes that have been measured in exclusive UPC events by ALICE using the central barrel detectors, namely $\gamma\gamma\rightarrow\mu\mu$, $\gamma\gamma\rightarrow KK$, $\jpsi\rightarrow\mu\mu$, $\jpsi\rightarrow \text{p}\bar{\text{p}}$, $\psip\rightarrow\mu\mu$, $\psip\rightarrow\jpsi\pi\pi$. This sample is referred to below as the \textit{training sample without exotica}. 

Each event includes the momenta \((p_x, p_y, p_z)\) and energy \(E\) of each particle, as well as the total invariant mass of the system. The $dE/dx$ value for each particle is obtained by requiring that particles of a given species and momentum follow a Gaussian distribution and fall within the 3 $N\sigma$ band shown in Fig.~\ref{fig:pid} for that particle species. 
%When the functional form of the TPC PID is applied, the $dE/dx$ value is obtained, as described above. 
Since experimentally the particle species would not be known, we use the generated $dE/dx$ value to also compute the corresponding $N\sigma_i$ values for other particle species, where $i$ indicates the particle species. 
%the hypothesis is related to. 
The structure of the events generated is shown in Tab.~\ref{tab:generation}, complete with kinematics and energy loss-related observables. Each line corresponds to a different track, and tracks belonging to the same event feature the same event identifier \verb|EvID|. 
\begin{table}[ht!]
    \centering
    \footnotesize
    \begin{tabular}{|c|c|c|c|c|c||c|c|c|c|c|}
    \hline
    \multicolumn{6}{|c||}{Kinematics} & \multicolumn{5}{c|}{PID Quantities} \\
    \hline
    EvID & Particle & $p_x$ & $p_y$ & $p_z$ & $E$ & $N\sigma_{e}$ & $N\sigma_{\mu}$ & $N\sigma_{\pi}$ & $N\sigma_{K}$ & $N\sigma_{p}$ \\
    \hline
    0 & muon & 0.1566 & -1.007 & -1.145 & 1.536 & -2.402 & -0.1276 & -0.0684 & 0.2207 & -0.5063 \\
    \hline
    0 & muon & -0.1566 & 1.007 & 1.145 & 1.536 & -2.381 & -0.0535 & 0.007124 & 0.3029 & -0.4409 \\
    \hline
    1 & muon & -0.00115 & -1.517 & -0.2142 & 1.536 & -2.371 & -0.0200 & 0.04116 & 0.3396 & -0.4123 \\
    \hline
    1 & muon & 0.00115 & 1.517 & 0.2142 & 1.536 & -2.4567 & -0.3150 & -0.2593 & 0.01262 & -0.6723 \\
    \hline
    2 & muon & -0.01927 & -0.06291 & -1.606 & 1.611 & -2.449 & -0.2902 & -0.2332 & 0.06856 & -0.5802 \\
    \hline
    \end{tabular}
    \caption{Sample of a track-level table obtained for the process $\jpsi \rightarrow \mu^+\mu^-$.}
    \label{tab:generation}
\end{table}

In the first implementation, we use the following strategy. First, all the tracks in Tab.~\ref{tab:generation} are grouped using their event identifier. For each track, the PID values $N\sigma_i$, where $i = e, \mu, \pi, K, $ p, are compared to a reference value, $N\sigma_{ref}$. To replicate typical experimental selections, the reference value could be taken to be $N\sigma_{ref}$ = 0.5, 1, or 2. If the value of $N\sigma_i$ is less than the reference, the track is considered a candidate of that particle type. We see roughly similar results for all choices of $N\sigma_{ref}$; in the following, we have shown the results obtained with $N\sigma_{ref}$ = 0.5. The track can be considered a candidate for more than one particle type. This is done on purpose, to be able to handle cases such as misidentification, or statistical fluctuations on the energy loss. The input vector for this autoencoder combines the information from each track to provide event-level information about the number of tracks in the event, and how many tracks can be considered as candidates for each particle type. An example of the input vectors for the first autoencoder, produced from simulated $\jpsi \rightarrow \mu^+\mu^-$ data, is shown in Tab.~\ref{tab:autoencoder-1}.

\begin{table}[ht!]
    \centering
    \begin{tabular}{|c|c|c|c|c|c|c|}
    \hline
         EvID  & N Electron & N Muon & N Pion & N Kaon & N Proton & N Tracks\\
           & candidates & candidates & candidates & candidates & candidates & \\
         \hline
         0 & 0 & 2 & 2 & 2 & 1 & 2\\
         \hline
         1 & 0 & 2 & 2 & 2 & 1 & 2\\
         \hline
         2 & 0 & 2 & 1 & 1 & 1 & 2\\
         \hline
         3 & 0 & 2 & 2 & 2 & 1 & 2\\
         \hline
         4 & 0 & 1 & 2 & 2 & 1 & 2\\
         \hline
    \end{tabular}
    \caption{Event-based tables for autoencoder option 1, using $\jpsi \rightarrow \mu^+\mu^-$ simulated data, from Tab.~\ref{tab:generation}.}
    \label{tab:autoencoder-1} 
\end{table}

In the second autoencoder implementation, rather than making a binary decision about whether a track is a candidate for a particular species, we have built input vectors using the actual $N\sigma_i$ values for each particle in the sample. Since the processes considered here produce a maximum of 4 tracks in the final state and there are 5 possible particle types produced, each line in the event record contains the total number of tracks in the event and 20 different values for N$\sigma$. To be able to handle processes with fewer tracks using the same structure for input vectors, we fill the unneeded cells with very large $N\sigma_i$ values. 
% An example of how a single event is rendered as input vector to this second autoencoder is shown in Tab.~\ref{tab:autoencoder-2}, where, due to space limitations, a single vector is displayed in four different rows, where the event identifier and the number of tracks is repeated in each row of the displayed table for clarity.
% \begin{table}[ht!]
%     \centering
%     \begin{tabular}{|c|c|c|c|c|c|c|}
%     \hline
%          Event ID $= j$ & Tracks & $N\sigma_{e}^1$ & $N\sigma_{\mu}^1$ & $N\sigma_{\pi}^1$ & $N\sigma_{K}^1$ & $N\sigma_{\text{p}}^1$\\
%          Event ID $= j$ & Tracks & $N\sigma_{e}^2$ & $N\sigma_{\mu}^2$ & $N\sigma_{\pi}^2$ & $N\sigma_{K}^2$ & $N\sigma_{\text{p}}^2$\\
%          Event ID $= j$ & Tracks & $N\sigma_{e}^3$ & $N\sigma_{\mu}^3$ & $N\sigma_{\pi}^3$ & $N\sigma_{K}^3$ & $N\sigma_{\text{p}}^3$\\
%          Event ID $= j$ & Tracks & $N\sigma_{e}^4$ & $N\sigma_{\mu}^4$ & $N\sigma_{\pi}^4$ & $N\sigma_{K}^4$ & $N\sigma_{\text{p}}^4$\\         
%          \hline
%     \end{tabular}
%     \caption{Event-based table for the second autoencoder. Owing to space limitations, a single event had to split in four rows, only for the purpose of explaining the structure used in this case. } {\bf there aren't any values in this table?  I assume there should be?}
%     \label{tab:autoencoder-2}
% \end{table}

In both cases, the event identifier column is stripped off the input vector before the training. Both autoencoders feature an encoder stage made of a single fully connected dense layer - each neuron is connected to the next layer - using a Rectified Linear Unit (ReLU) activation function \cite{dertat}, and the decoder stage is similarly made of a single dense layer with a sigmoid activation function instead. The input data are transformed using a \verb|MinMax| scaler \cite{scikit-learn}.

Both autoencoders are trained using the \textit{training sample without exotica} sample, with the loss function of choice being the mean squared error (MSE), which is measured using the squared differences between the input and reconstructed output:
\[
\text{MSE} = \frac{1}{\sigma} \sum_{i=1}^{\sigma} (x_i - \hat{x}_i)^2
\]
where $\sigma$ is the number of features defined above, $x$ is the input and $\hat{x}$ is the reconstructed feature vector. We have implemented early stopping by halting the iterative procedure of the training if the performances did not improve after ten iterations of the training. 

\begin{figure}[ht!]
	\begin{center}
		\subfigure[]{%Caption of Second Figure]{%
			\label{fig:loss}
			\includegraphics[width=0.45\textwidth]{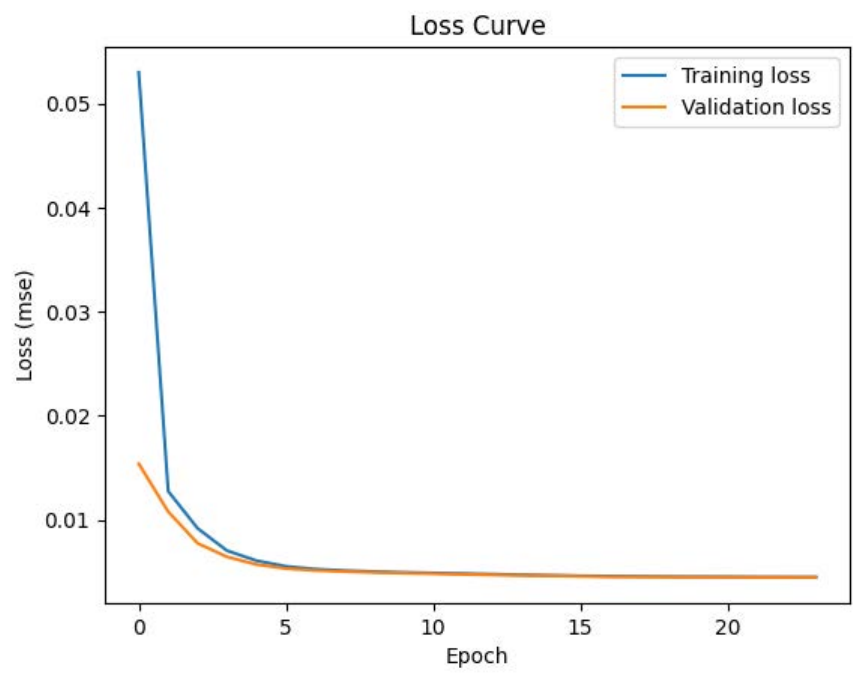}
		}
            \subfigure[]{%Caption of First Figure]{%
			\label{fig:loss-2}
			\includegraphics[width=0.5\textwidth]{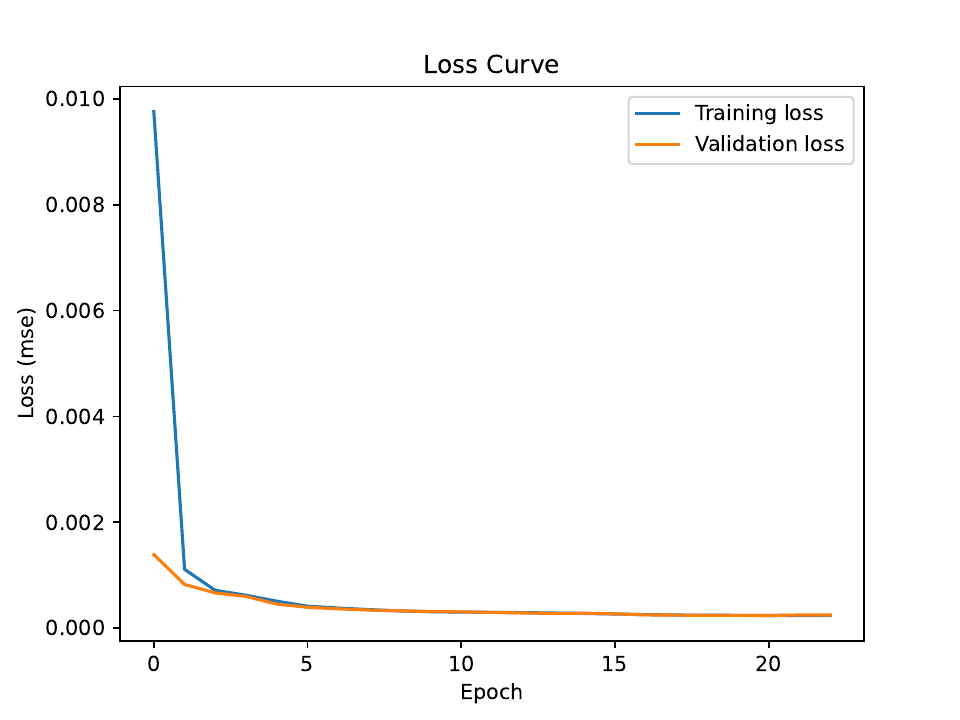}
		}\\ %  ------- End of the first row ----------------------%
	\end{center}
	\caption{Fig.~\ref{fig:loss} and \ref{fig:loss-2} show the loss curves for the first and second autoencoder implementations, respectively.}
	\label{fig:losses}
\end{figure}

% The typical loss curve for the training of the autoencoders is shown in Fig.~\ref{fig:loss}.
The typical loss curves for the training of the autoencoders is shown in Fig.~\ref{fig:loss} and \ref{fig:loss-2}.
The training and validation loss curves both drop sharply in the first few epochs and then level off below 0.005 for both autoencoders, indicating that they have captured the bulk of the underlying data structure. The very good overlap of the two curves and the use of early stopping at the plateau show that the models generalize well without overfitting. 
% \begin{figure}[b]
% \centering
% \includegraphics[width=0.6\columnwidth]{fig/loss_curve.pdf}
% \caption{\label{fig:loss} Loss curve for the first autoencoder implementation. } 
% \end{figure}

% In the following, the discussion will focus on the first design of the autoencoder, i.e. Tab.~\ref{tab:autoencoder-1}, but similar results are achieved with the second autoencoder. 

\subsection{Generating a realistic particle cocktail}
\label{sec:eficiency}
All the processes shown in Fig.~\ref{fig:tpc-response} have been measured by the ALICE Collaboration.  We have built a particle cocktail of these processes that is meant to replicate a realistic data set collected by the ALICE detector. This baseline data set can be injected with additional exotic processes to ascertain the effectiveness of anomaly detection-based techniques. 

The starting point is the extensive literature produced by the ALICE collaboration using the Run 2 data set. The 
%final independent 
particle cocktails used to test the design of the autoencoders are built starting from the number of events measured for each process by ALICE.
For the resonances, this is quite straightforward, since \cite{Acharya:2748581} shows the number of events measured by ALICE for $\jpsi\rightarrow\mu\mu$, $\jpsi\rightarrow \text{p}\bar{\text{p}}$, $\psip\rightarrow\mu\mu$, and for $\psip\rightarrow\jpsi\pi\pi$ with Run 2 luminosity. 
%In the case of the resonances, this is quite straightforward, since \cite{Acharya:2748581} shows that ALICE has measured $\jpsi\rightarrow\mu\mu$ with 3,120 events, $\jpsi\rightarrow \text{p}\bar{\text{p}}$ with 61 events, $\psip\rightarrow\mu\mu$ with 58 events, and $\psip\rightarrow\jpsi\pi\pi$ with 53 events. 

The continuum distributions require an ad hoc treatment. To determine how many $\gamma\gamma \rightarrow K^+K^-$ events to add to the cocktail, we begin with the equation for the differential cross section \cite{ALICE:2023kgv}: 
\begin{equation}
\frac{d^{2}\sigma}{dM_{KK}dy_{KK}} = \frac{N_{KK} \times f_{pileup}}{\AxE \times \mathscr{L} \times \Delta M_{KK} \times \Delta y_{KK}}\text{ ,}
\label{eq:sigma-kk}
\end{equation}
where $M_{KK}$ represents the invariant mass of the dikaon pairs, $y_{KK}$ is their rapidity, $N_{KK}$ the number of events per mass bin, $f_{pileup}$ the pile-up, $\AxE$\ the acceptance-cross-efficiency, $\mathscr{L}$ the integrated luminosity of the sample, $ \Delta M_{KK}$ the size of the invariant mass bin the data point corresponds to, and $\Delta y_{KK}$ the same for the rapidity bin. The meaning of the variables is also given in \cite{ALICE:2023kgv}.

As \cite{ALICE:2023kgv} does not report the number of events per invariant mass bin, nor each of the values used in Eq.~\ref{eq:sigma-kk}, we infer the number of events from the statistical uncertainties. It is possible to express Eq.~\ref{eq:sigma-kk} as:
\begin{equation}
\frac{d^{2}\sigma}{dM_{KK}dy_{KK}} = \alpha \cdot N_{KK}\text{ .}
\label{eq:sigma-kk-2}
\end{equation}
We assume that the statistical uncertainty in the cross section is driven by the uncertainty in the number of events in each invariant mass bin in \cite{ALICE:2023kgv}. This is quite reasonable, since the statistical uncertainty has two contributions, one from the \AxE\ correction, and the other from the number of events. The former originates from plentiful official ALICE simulations and is quite small compared to the latter. This assumption implies that:   
\begin{equation} \label{eq:kaon}
\frac{\Delta \sigma}{\sigma} = \frac{\alpha \sqrt{N_{KK}}}{\alpha N_{KK}} = \frac{1}{\sqrt{N_{KK}}}\text{ ,}
\end{equation}
where $\sigma$ and $\Delta \sigma$ are the cross section values and their uncertainties.  Using the measured values in \cite{ALICE:2023kgv} for the mass range of 1.1 to 1.4~\GeVmass, we obtain a value of $N_{KK} = 34$ events within this small mass range. We use this value to normalize the exponential distribution. By integrating the normalized distribution, we determine the number of events 
%shown in Tab.~\ref{table:combined_runs} 
in the mass region above $2m_K$, where $m_K$ is the kaon mass. 

The process to determine the number of $\gamma\gamma \rightarrow \mu^+\mu^-$ continuum events to inject into our cocktail is similar to that of the $\gamma\gamma \rightarrow K^+K^-$ events. The slope of the exponential was obtained by fitting the invariant mass distributions in \cite{Acharya:2748581} to an exponential. 
%Following similar considerations as above for the normalisation of the integral, we introduce a sample of 169,203 dimuon continuum events. {\bf this needs more explanation}

We combine the calculated numbers of events from the different processes into one data set that approximately matches what is observed.
%in ALICE Run 2 data. 
This will be referred to in the following as the \textit{no exotica} sample. Integrated luminosity values for Run 2, Run 3, and Runs 3 and 4 combined are shown in Tab.~\ref{table:combined_runs}, along with the number of events from each process included in the \textit{no exotica} sample.

It is then possible to introduce events from new processes. We first inject a number of $\jpsi \rightarrow 4\pi$ events. The number of injected events is determined using:
\begin{equation}
N_{\jpsi\rightarrow 4\pi} = \sigma \times BR\times \mathscr{L} \times A \times \epsilon \text{ ,}
\end{equation}
where $N_{\jpsi\rightarrow 4\pi}$ is the number of $\jpsi \rightarrow 4\pi$ events, $BR$ the branching ratio,  $\sigma$ the cross section, $\mathscr{L}$ the integrated luminosity of the sample, and $A \times \epsilon$ the acceptance $\times$
efficiency.
The branching ratio is found from the PDG \cite{ParticleDataGroup:2024cfk}.  ALICE has measured the production cross section for coherent \jpsi in \cite{Acharya:2748581} as the average of the cross sections obtained from three different channels, namely $\jpsi \rightarrow ee$, $\jpsi \rightarrow \mu\mu$ and $\jpsi \rightarrow \text{p} \bar{\text{p}}$, which we assume as the cross section for this process as well. ALICE has also published the efficiency for the measurement of $4\pi$ events in the central barrel in \cite{ALICE:2024kjy}, where the efficiency plateaus at higher masses to an efficiency value of 0.07 which we consider as the basis of our estimate. Tab.~\ref{table:combined_runs} shows the number of $\jpsi \rightarrow 4\pi$ events expected in Run 2, Run 3, and Runs 3 and 4 combined. 

% Using these values we find:

% \begin{equation}
% N_{J/\psi\rightarrow4\pi}=(4.1 \times 10^6 \text{ nb})(0.0036)(1 \text{ nb}^{-1})(0.07) = 947.1.
% \end{equation}

% This gives us a value of about 947 $J/\psi\rightarrow4\pi$ processes to inject into our Run 3 cocktail. Scaling down to Run 2 gives about 237 processes.
%-----------

We have also introduced a sample of $\pentaquark \rightarrow \jpsi + \text{p}$ events. We estimate the number of events to be injected starting from:
\begin{equation}
    N_{\pentaquark} = \sigma \times BR\times \mathscr{L} \times A \times \epsilon\text{ ,}
\end{equation}
where $N_{\pentaquark}$ is the number of pentaquarks, $BR\times \sigma$ the branching ratio times cross section, $\mathscr{L}$ the integrated luminosity of the sample, and $A \times \epsilon$ the acceptance $\times$
efficiency.
Based on the upper limits for the branching ratio, $BR$, and that of the $BR$ times cross section, $BR\times \sigma$, measured by GlueX \cite{GlueX:2019mkq}, which are 4.6\% and 4.6~nb, respectively, we derive  a cross section of 100~$\text{nb}$. 
%As we are interested in providing a realistic ballpark estimate of the number of events to be introduced, we commence then from a baseline of 100~$\text{nb}$ for the cross section, by dividing these values. 
\cite{GlueX:2019mkq} provides several possible branching ratios: 0.14\%, 4.6\%, and 79\%, depending on the production mechanism.  
%The integrated luminosity for the 2018 sample used in \cite{Acharya:2748581} by ALICE is about 250$ \text{ }\mu \text{b}^{-1}$. We therefore calculate:
% \begin{equation}
% N_{\pentaquark} = (0.1\text{ }\mu\text{b})  
% \begin{cases}
%    0.0014 \\
%    0.046 \\
%    0.79 \\
%  \end{cases}
%  \times (250 \text{ }\mu \text{b}^{-1})(\epsilon) = 
%  \begin{cases}
%  0.035 \\
%  1.15 \\
%  19.75 \\
%  \end{cases} \times (\epsilon),
% \end{equation}
%which shows that the number of pentaquarks observed could range from zero to twenty, depending on the experimental efficiency. To first approximation, 
%which is realistic due to the low numbers involved with the pentaquark production, 
We take the efficiency to be the product of the efficiency for \jpsi and for a proton. We use the efficiency values found in \cite{ALICE:2021dtt} for \jpsi in the momentum range of 0.7 to 0.8~\GeVc which is close to the kinematically allowed range for the \jpsi produced from the decay of the pentaquark, with an efficiency between about 0.076 and 0.16. The TPC efficiency for protons from \cite{ragoni2018hadron} is about 0.76. We then obtain an upper limit for the detection efficiency of the \pentaquark of 12.16\% through the muon channel. We factor in the detection of the \jpsi though additional channels, such the $ee$, $4\pi$ and $2\pi2K$ channel, leading to about a factor 2.5 increase in overall efficiency compared to the muon channel alone. Ultimately, Tab.~\ref{table:combined_runs} shows the number of $\pentaquark \rightarrow \jpsi + \text{p}$ events expected in all $\jpsi$ decay channels in Run 2, Run 3, and Runs 3 and 4 combined, obtained with the highest of the branching ratio scenarios provided in \cite{GlueX:2019mkq}.

%As we are interested in the decay channel of $\jpsi\rightarrow\mu\mu$ even for the decay of the \pentaquark, and since due to kinematic constraints, the 
% Since the kinematically allowed region for the muons produced through $\pentaquark\rightarrow\jpsi + \text{p}\rightarrow\mu\mu + \text{p}$ is very close to that of the muons from coherently photoproduced \jpsi, we take the efficiency value from \cite{Acharya:2748581} directly, i.e. $\epsilon = 0.037$. The kinematically allowed region for the proton is rather similar to that of the protons of $\jpsi\rightarrow \text{p}\bar{\text{p}}$, so we take the efficiency for this channel from \cite{Acharya:2748581},  $\epsilon = 0.023$. We make the further approximation that the efficiency of a single proton is $\epsilon_{\rm p}  = \sqrt{\epsilon_{\jpsi\rightarrow \text{p}\bar{\text{p}}}}$. 
%To account for the decay of the pentaquark only including one proton, the square root is taken of the proton's efficiency value.

% These values lead to a detection efficiency of 0.56\% for $\pentaquark\rightarrow\jpsi + \text{p}\rightarrow\mu\mu + \text{p}$ . Tab.~\ref{table:combined_runs} shows the number of $\pentaquark \rightarrow \jpsi + \text{p}$ events expected in all $\jpsi$ decay channels in Run 2, Run 3 and Runs 3 and 4 combined.

%--------

\begin{figure}[b]
\includegraphics[width=1.\columnwidth]{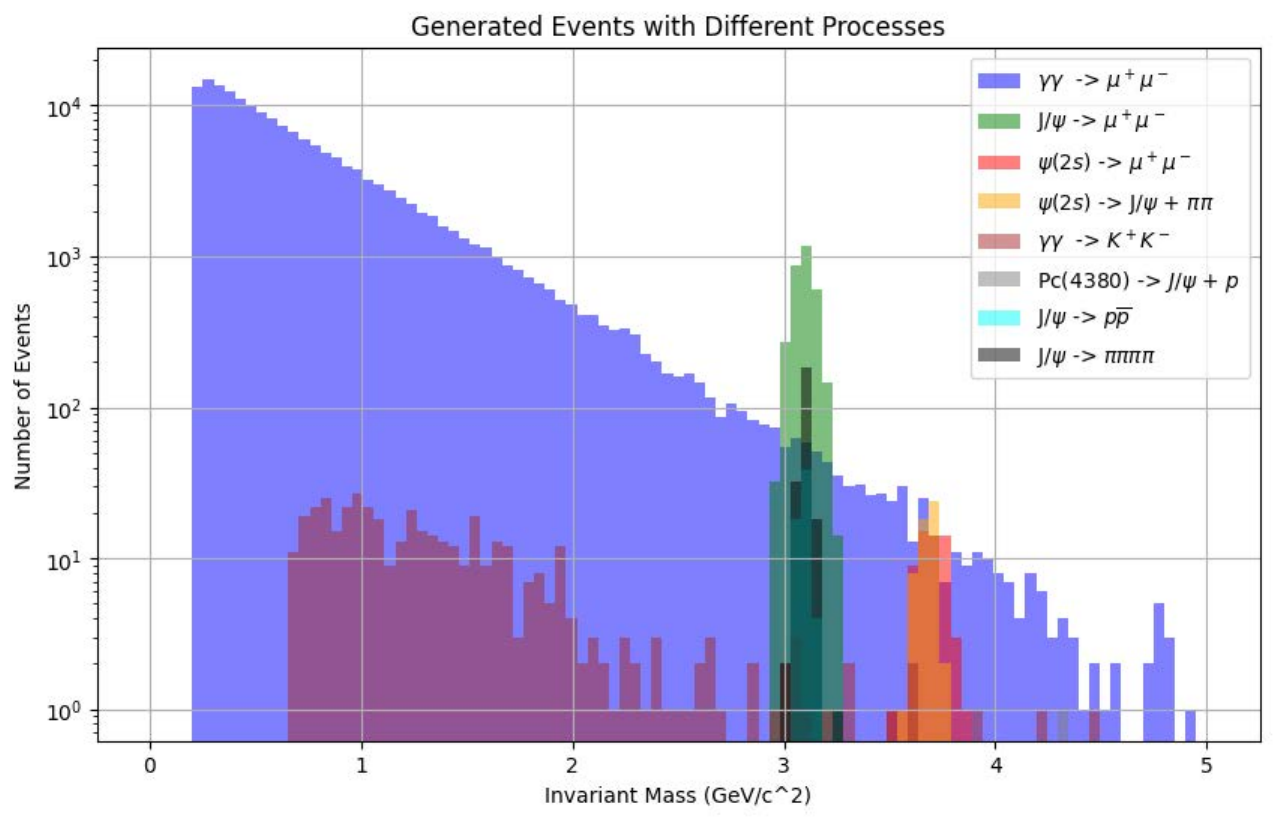}
\caption{\label{fig:cocktail} Invariant mass distribution of the \textit{with exotica} particle cocktail representing the processes measured by the ALICE Collaboration using the Run 2 data set, for masses above 1.5~\GeVmass as well as the exotic processes injected in the cocktail, $\jpsi\rightarrow4\pi$ and $\pentaquark\rightarrow\jpsi+\text{p}$.} 
\end{figure}

The \textit{no exotica} sample  with the addition of \pentaquark and $\jpsi \rightarrow 4\pi$ events will be referred to below as the \textit{with exotica} sample. The invariant mass distribution generated for the \textit{with exotica} sample using Run 2 luminosity, is shown in Fig.~\ref{fig:cocktail}, with different particle species represented in different colours.

%---

\begin{table}[]
\centering
\begin{tabular}{|l|r|r|r|}
\hline
Decay channel & Run 2 (250 $\mu b^{-1}$) & Run 3 (1 $nb^{-1}$) & Run 3 \& 4 (10 $nb^{-1}$) \\ \hline
$\gamma\gamma \rightarrow \mu\mu$ & 169,203 & 676,812 & 6,768,120 \\ \hline
$\gamma\gamma \rightarrow KK$     & 423     & 1,692   & 16,920    \\ \hline
$\jpsi\rightarrow\mu\mu$         & 3,120   & 12,480  & 124,800   \\ \hline
$\jpsi\rightarrow \text{p}\bar{\text{p}}$      & 61      & 244     & 2,440     \\ \hline
$\psip\rightarrow\mu\mu$          & 58      & 232     & 2,320     \\ \hline
$\psip\rightarrow \jpsi\pi\pi$   & 53      & 212     & 2,120     \\ \hline
$\jpsi\rightarrow4\pi$           & 237     & 947     & 9,470     \\ \hline
$\pentaquark\rightarrow \jpsi+\text{p}$   & 0 to 1  & 1 to 2  & 14        \\ \hline
% $\pentaquark\rightarrow \jpsi+\text{p}$   & 0.3554  & 1.4216  & 14        \\ \hline

\end{tabular}
\caption{Number of events for each decay channel across different luminosities.}
\label{table:combined_runs}
\end{table}

\subsection{Testing the Autoencoders}

\begin{figure}[ht!]
	\begin{center}
		\subfigure[]{%Caption of Second Figure]{%
			\label{fig:mse}
			\includegraphics[width=0.95\textwidth]{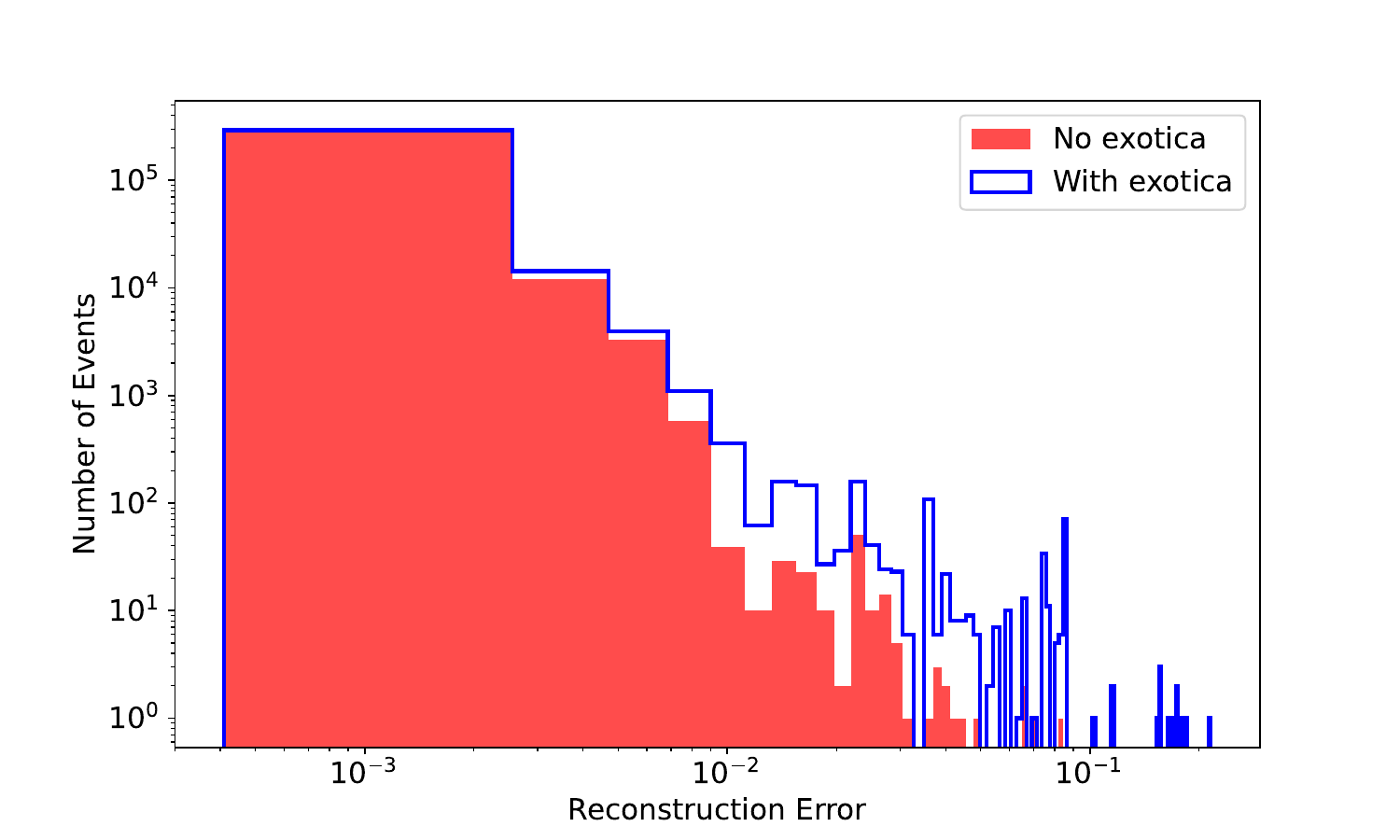}
		}
            \subfigure[]{%Caption of First Figure]{%
			\label{fig:mse-2}
			\includegraphics[width=0.95\textwidth]{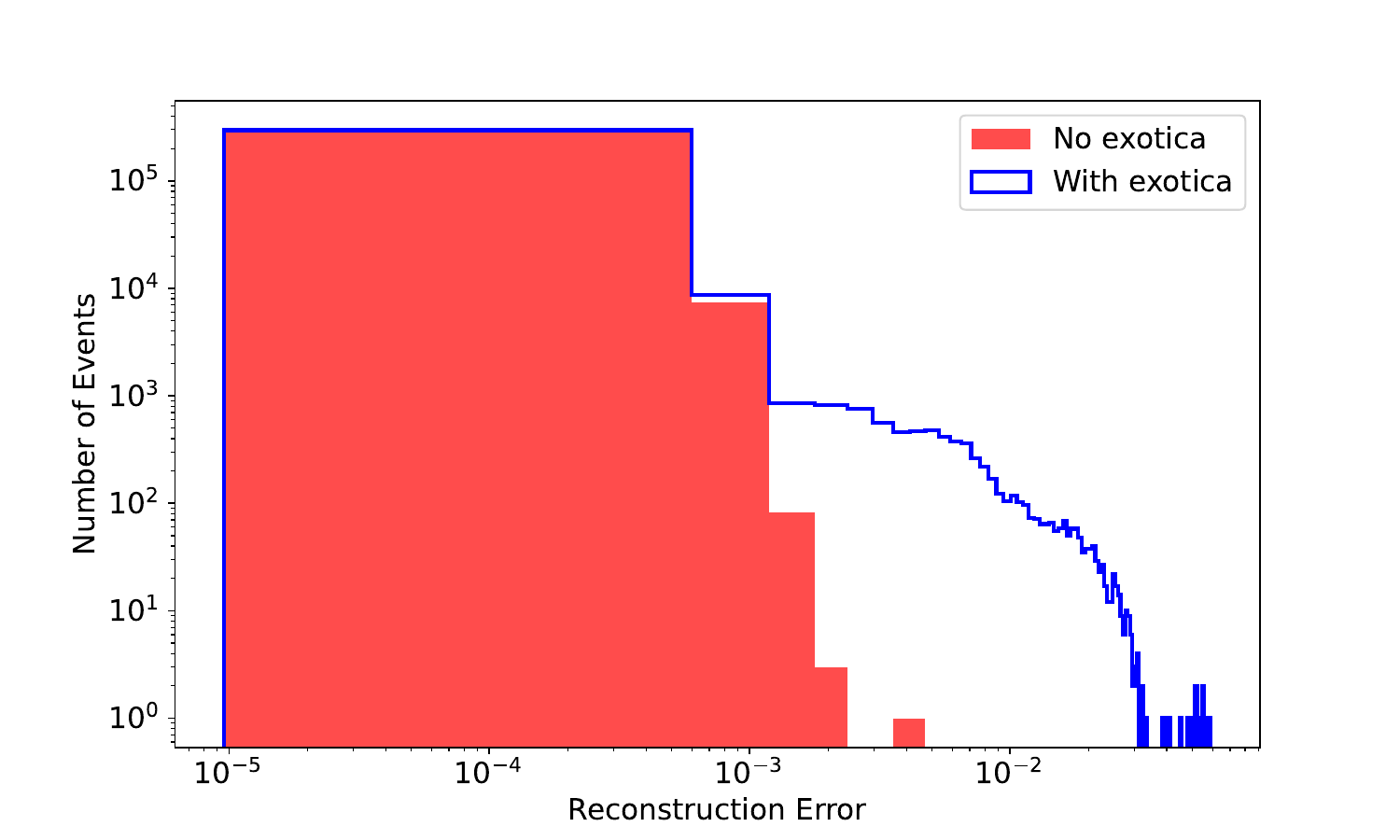}
		}\\ %  ------- End of the first row ----------------------%
	\end{center}
	\caption{Fig.~\ref{fig:mse} and \ref{fig:mse-2} show the distributions of the mean squared error (MSE) for the first and second autoencoders, respectively, for the two particle cocktails, one with only processes previously measured by ALICE, and the other with additional exotic processes. The latter can be flagged by this technique owing to high values of MSE. Variable binning is used to enhance the visibility of the few high MSE events.}
	\label{fig:mse-all}
\end{figure}

The two autoencoders are then applied to the two cocktails representing the ALICE data, the \textit{no exotica} and \textit{with exotica} samples.  The distribution of mean squared errors (MSE) for each cocktail is shown in Figure~\ref{fig:mse} and \ref{fig:mse-2} for the first and second autoencoder implementations, respectively. The \textit{with exotica} cocktail features higher values of mean squared error (MSE) than the \textit{no exotica} cocktail, which includes only processes the autoencoders were trained upon. Exotic processes can then be selected using a MSE selection.
% \begin{figure}[b]
% \includegraphics[width=1.\columnwidth]{fig/updated_Run4_error.pdf}
% \caption{\label{fig:mse} Distribution of the mean squared error (MSE) for the events of the two particle cocktails, one with only 
% processes previously measured by ALICE, and the other with additional exotic processes. The latter can be flagged by this technique owing to high values of MSE. Variable binning is used to enhance the visibility of the few high MSE events.}
% \end{figure}

\begin{figure}[ht!]
	\begin{center}
		\subfigure[]{%Caption of Second Figure]{%
			\label{fig:flagged}
			\includegraphics[width=0.95\textwidth]{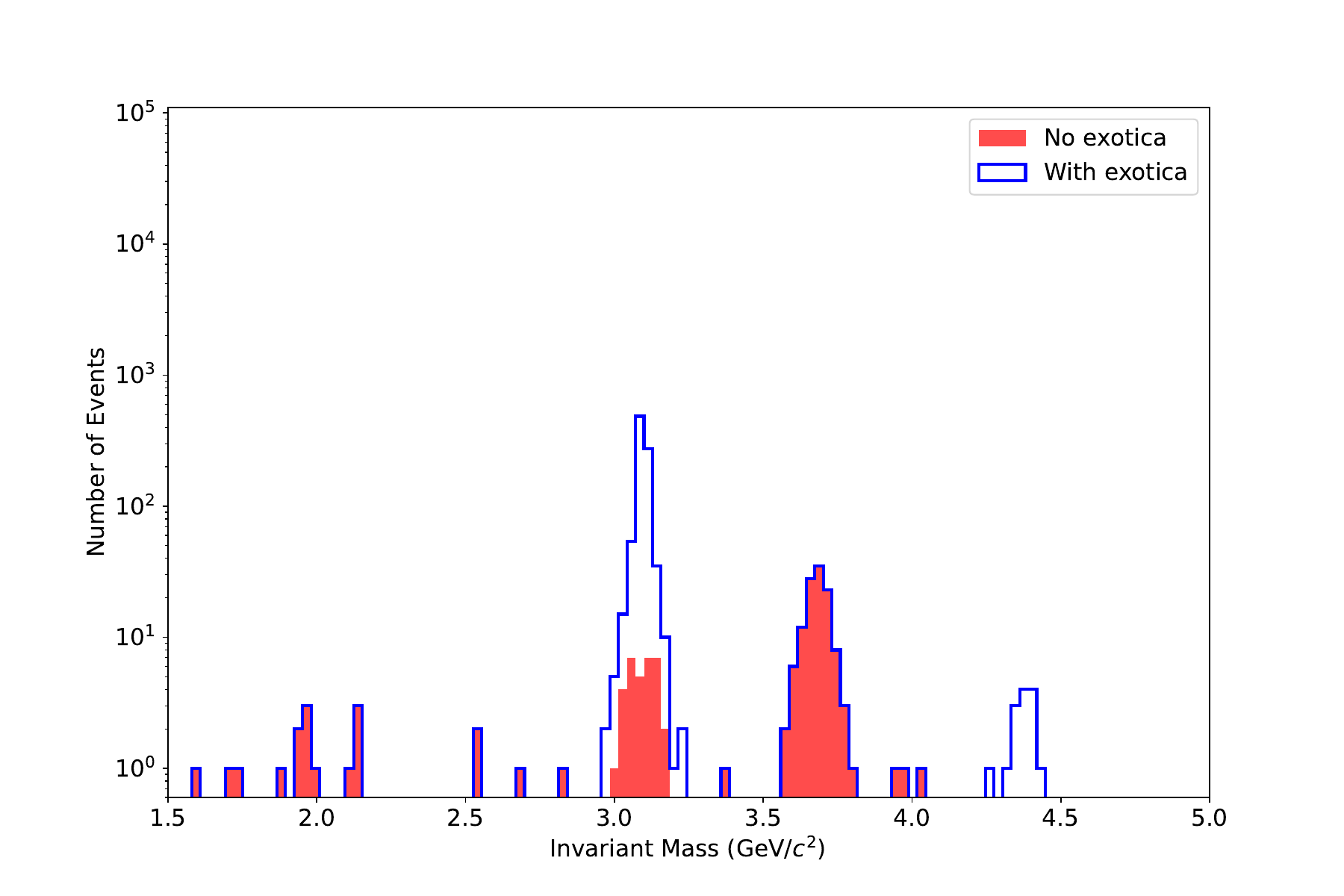}
		}
            \subfigure[]{%Caption of First Figure]{%
			\label{fig:flagged-2}
			\includegraphics[width=0.95\textwidth]{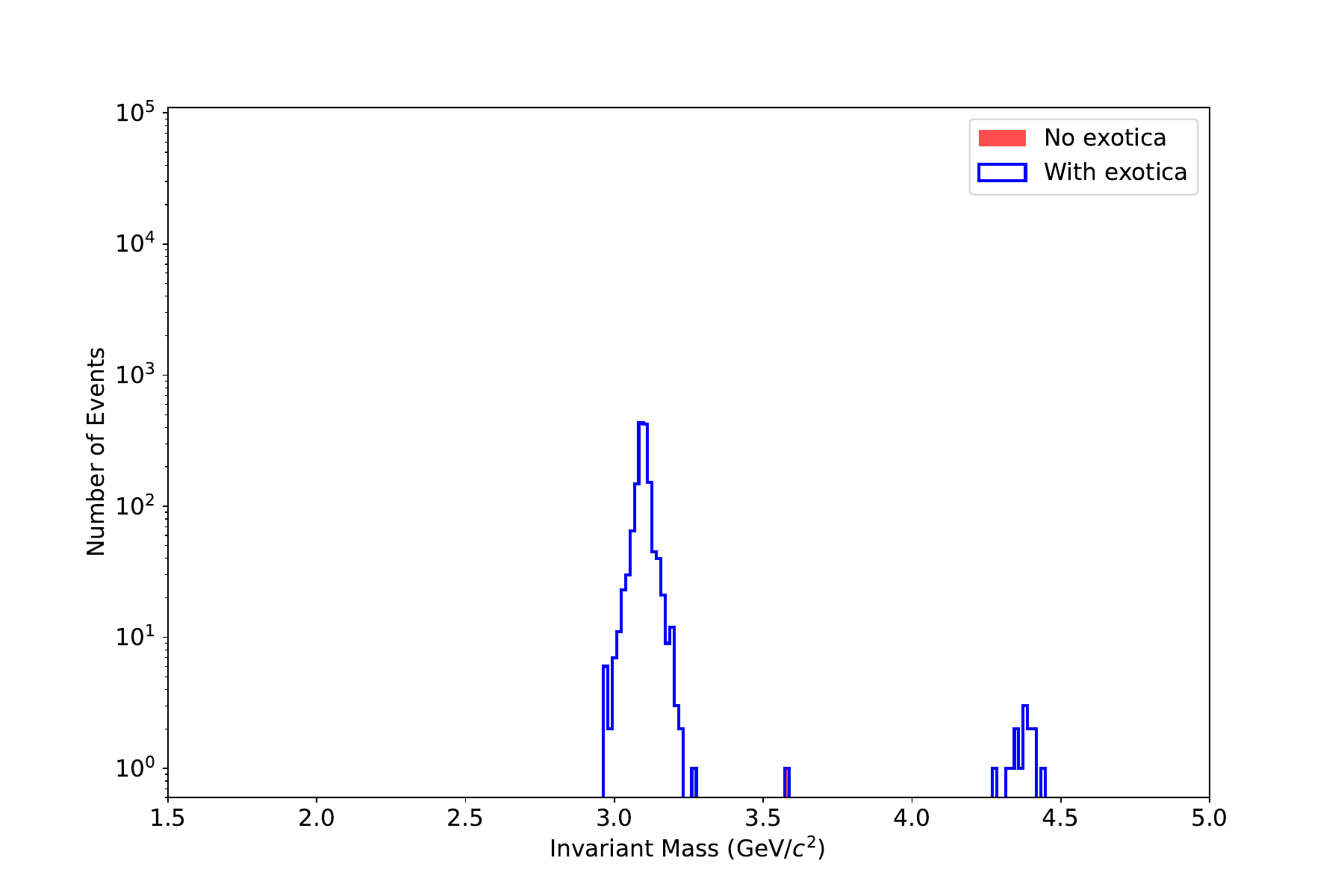}
		}\\ %  ------- End of the first row ----------------------%
	\end{center}
	\caption{Fig.~\ref{fig:flagged} and \ref{fig:flagged-2} show the invariant mass distribution of those events which were flagged by the autoencoder as anomalous, i.e. with MSE values above 0.01, for the first and second autoencoders, respectively. } 
	\label{fig:flagged-all}
\end{figure}

Events with MSE above 0.01, corresponding to the top 4$\%$ in the MSE distribution, were selected as potential exotica. The invariant mass distributions of the events selected as potential exotica are shown in Figure~\ref{fig:flagged} and \ref{fig:flagged-2} for the first and second autoencoders, respectively, separately for the \textit{no exotica} and \textit{with exotica} samples. Two peaks can be visually inspected at about 3.1~\GeVmass and 4.4~\GeVmass, corresponding to the injected sample of $\jpsi\rightarrow 4\pi$ and $\pentaquark\rightarrow\jpsi+\text{p}$. Two figures of merit may be employed: the purity, which is defined as the fraction of events flagged in a particle mass region coming from one of the decay channels of the particle rather than from a contribution of the continuum \cite{cowan1998statistical}, and the efficiency of a decay channel, which is the fraction of input events for that decay channel flagged by this technique. $\jpsi$s were selected with very high purity, with a contamination of less than 10\% due to other \jpsi processes using the first design, and no contamination at all using the second design. The efficiency of flagging the injected rare $\jpsi$ decay was about 11\%  for the first design and above 95\% for the second design. At most a few pentaquarks are expected to be measured, the estimates of which can be found in Tab.~\ref{table:combined_runs}. Both autoencoders flag all the injected pentaquark events as anomalous, with no contamination from other processes, as shown in Figure~\ref{fig:flagged} and \ref{fig:flagged-2}, demonstrating very high efficiency of the technique for pentaquark detection for both designs.
% \begin{figure}[bh!]
% \includegraphics[width=1.\columnwidth]{fig/updated_Run4_mass.pdf}
% \caption{\label{fig:flagged} Invariant mass distribution of those events which were flagged by the autoencoder as anomalous, i.e. with high MSE values.}
% \end{figure}

\clearpage
\section{Discussion}
These results show the impact that modern anomaly detection through machine learning might have on searches for rare processes and for new physics in low-background events such as those typical of exclusive physics, i.e. diffractive and UPC events. Anomaly detection is quite suitable in searches for new physics since it is possible to train a model solely on previously measured processes, without making any assumptions about the new process to be investigated.

As described in \cite{Ragoni:2024ovv}, this approach can be used to complement more traditional searches. For example, anomaly detection can be used to isolate interesting events with high purity, without having to apply kinematic selections or anything which can further bias the sample. Once the event topology for these events is understood, a traditional measurement can be performed, this time to achieve higher statistics rather than purity. In \cite{Ragoni:2024ovv}, it is also noted how autoencoders operate at much higher dimensionalities compared to more traditional approaches, which usually involve looking at only \pt, $y$ and the invariant mass of the system.

In this application, autoencoders were selected over more sophisticated Variational Autoencoders (VAE) for two reasons. First, the autoencoders already show excellent performance in terms of purity and efficiency of pentaquarks selected despite their simple design. Second, typical computing requirements for the deployment of machine learning models often leverage conversions through ONNX \cite{ONNX:2021}. In this case, autoencoders are preferable to VAEs because they can be converted naturally within this framework.

% Due to outstanding performances in terms of purity and efficiency of pentaquark detection of such a simple design, and owing to the computing requirements of the analysis framework of experiments such as ALICE, an autoencoder is preferable to a Variational Autoencoder (VAE). In ALICE O2 \cite{AliceO2Group} in fact, machine learning models can only be ported through conversion with e.g. ONNX \cite{ONNX:2021}, and autoencoders can be converted naturally within this framework. {\bf .  The bit about ALICE O2 is very experiment specific.  Do the same things apply to other experiments?  I guess VAEs cannot be converted through ONNX?}

\begin{figure}[ht!]
	\begin{center}
		\subfigure[]{%Caption of Second Figure]{%
			\label{fig:penta}
			\includegraphics[width=0.45\textwidth]{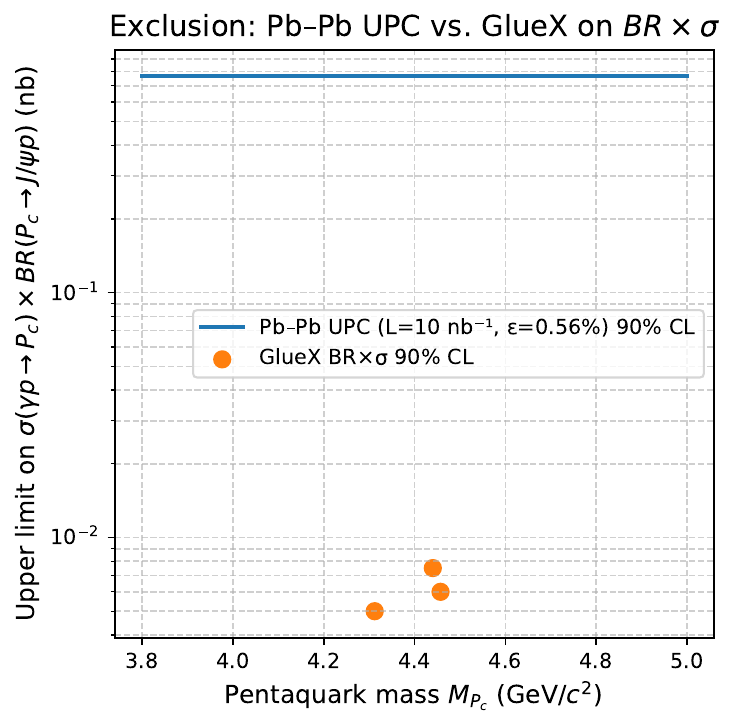}
		}
            \subfigure[]{%Caption of First Figure]{%
			\label{fig:tetra}
			\includegraphics[width=0.45\textwidth]{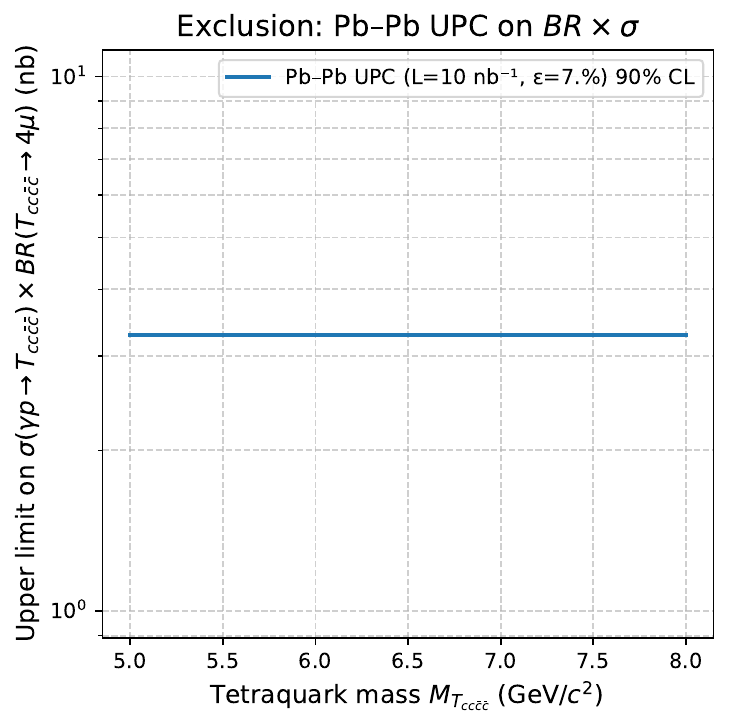}
		}\\ %  ------- End of the first row ----------------------%
	\end{center}
	\caption{Fig.~\ref{fig:penta} and \ref{fig:tetra} show the upper limits for the $\sigma\times BR$ for the discovery channels of the pentaquark $\pentaquark \rightarrow \jpsi+\text{p}$ and the fully-charmed tetraquark $\tetraquark \rightarrow 4\mu$. The upper limits are measured using UPC conditions, and the pentaquark upper limits are compared with those measured by GlueX.}
	\label{fig:limits}
\end{figure}

It is also possible to use a search based on anomaly detection to establish exclusion limits on the production of new resonances, or exotica, in UPCs and, more generally, in diffractive
events. We have computed an exclusion limit for the $\pentaquark$, shown in Figure ~\ref{fig:penta}. The limits are obtained considering the efficiency presented in Sec.~\ref{sec:eficiency}  for the pentaquark. The limits are simplified in this case, since the autoencoders are demonstrated to select these events with very high purity, so no background is being considered while measuring these limits. We consider the luminosity from the full Run 3 and 4 data sets. While the limit computed is well above the limit provided by GlueX \cite{GlueX:2019mkq}, this approach allows us to place a limit in a broad range of masses.

In addition, we compute an exclusion limit for the production of fully-charmed tetraquarks, shown in Figure ~\ref{fig:tetra}.  
For $\tetraquark\rightarrow 4\mu$ we use the efficiency that was used for the estimates of the $\jpsi\rightarrow 4\pi$ process, since the muons at those energies will feature similar PID signals to the pions, and are expected to show similar efficiencies.  To the best of our knowledge, no upper limit of tetraquark production in UPC events has been measured, so a UPC measurement would provide the state-of-the-art for masses above 5~\GeVmass. 

\clearpage
\section{Summary}
The results shown here demonstrate the capabilities of machine learning techniques to select with high purity exotic processes and rare decays in samples of exclusive events. This technique can be used to perform a first high-purity measurement, to infer the topology of the rare process under investigation, before performing a more traditional search in invariant mass. It is also possible to use a search based on anomaly detection to establish exclusion limits on the production of new resonances, or exotica, in UPCs and more generally, in diffractive events.

\acknowledgments

This work was funded by the Department of Energy (DOE) of the United States of America (USA) through the grant DE-FG02-96ER40991.

% Bibliography

% [A] Recommended: using JHEP.bst file
\bibliographystyle{JHEP}
\bibliography{biblio.bib}

%% or
%% [B] Manual formatting (see below)
%% (i) We suggest to always provide author, title and journal data or doi:
%% in short all the informations that clearly identify a document.
%% (ii) please avoid comments such as "For a review'', "For some examples",
%% "and references therein" or move them in the text. In general, please leave only references in the bibliography and move all
%% accessory text in footnotes.
%% (iii) Also, please have only one work for each \bibitem.

% \begin{thebibliography}{99}

% \bibitem{a}
% Author,
% \emph{Title},
% \emph{J. Abbrev.} {\bf vol} (year) pg.

% \bibitem{b}
% Author,
% \emph{Title},
% arxiv:1234.5678.

% \bibitem{c}
% Author,
% \emph{Title},
% Publisher (year).

% \end{thebibliography}
\end{document}